**Resonant Kuiper Belt Objects - a Review**

Renu Malhotra
Lunar and Planetary Laboratory, The University of Arizona, Tucson, AZ, USA
Email: renu@lpl.arizona.edu

## Abstract

Our understanding of the history of the solar system has undergone a revolution in recent years, owing to new theoretical insights into the origin of Pluto and the discovery of the Kuiper belt and its rich dynamical structure. The emerging picture of dramatic orbital migration of the planets driven by interaction with the primordial Kuiper belt is thought to have produced the final solar system architecture that we live in today. This paper gives a brief summary of this new view of our solar system's history, and reviews the astronomical evidence in the resonant populations of the Kuiper belt.

## Introduction

Lying at the edge of the visible solar system, observational confirmation of the existence of the Kuiper belt came approximately a quarter-century ago with the discovery of the distant minor planet (15760) Albion (formerly 1992 QB1, Jewitt & Luu 1993). With the clarity of hindsight, we now recognize that Pluto was the first discovered member of the Kuiper belt. The current census of the Kuiper belt includes more than 2000 minor planets at heliocentric distances between ~30 au and ~50 au. Their orbital distribution reveals a rich dynamical structure shaped by the gravitational perturbations of the giant planets, particularly Neptune.

Theoretical analysis of these structures has revealed a remarkable dynamic history of the solar system. The story is as follows (see Fernandez & Ip 1984, Malhotra 1993, Malhotra 1995, Fernandez & Ip 1996, and many subsequent works). Some ~4 gigayears ago, the orbits of the giant planets - Jupiter, Saturn, Uranus and Neptune - were more compact and the solar system contained a lot more planetary debris in the form of asteroids and comets. That debris was gradually cleared by the collective gravitational perturbations of the planets, but this process had a back-reaction on the planets: it caused a spreading out and re-arrangement of the giant planets' orbits and eventually led to a more stable solar system that we enjoy now.

During that epoch of planet migration, the populations of minor planets were decimated. The small fraction that survived in proximity to their formation locations are predominantly beyond Neptune. An early specific theoretical prediction was that the minor planets beyond Neptune that survived the planet migration epoch should be found piled up in eccentric orbits in mean motion resonances (MMRs) with Neptune, particularly the 3/2 and the 2/1 MMRs (Malhotra 1995). The subsequent discoveries of dozens of "Plutinos" in the 3/2 MMR has been interpreted as proof of the theory and has led to the widespread acceptance of the idea of giant planet migration as a core part of the early history of the solar system.

The migration of the giant planets has implications for a broad range of planetary science, including our understanding of planet formation, the origin of the solar system architecture, dynamical evolution of the solar system, the provenance of the various minor planet groupings, the transport of planetesimals throughout the solar system, the time history of the meteoroidal impact flux on the Earth and the Moon and on all the other planets and moons, the formation of the Oort Cloud of comets, and the history of ejection of planets and planetesimals from the solar system. The primary evidence for giant planet migration lies in the resonant populations of the Kuiper belt; we review this evidence here.

**Pluto as the first resonant Kuiper belt object**

Shortly after its discovery in 1930, it became obvious that Pluto was a very peculiar planetary object: much smaller than the giant outer planets, even smaller than the terrestrial planets, its orbit did not follow the pattern of nearly co-planar and nearly circular planetary orbits well-separated from each other. Pluto's mass is less than 20% that of the Moon, and comparable to the mass of the largest minor planet in the asteroid belt, Ceres. Its orbital plane is tilted ~17 degrees to the ecliptic. While its average orbital radius exceeds Neptune's by nearly 10 au, its elliptical orbit has perihelion distance less than Neptune's, making its orbit not well-separated from that planet. With numerical analysis of its long-term orbital motion, it was found that Pluto reaches perihelion at a longitude always well away from Neptune and always at a point well above Neptune's orbit plane. Although its perihelion distance is interior to Neptune's, the location of the perihelion librates with a period of about 20,000 years around a longitude separated from Neptune by 90 degrees; the longitude libration amplitude is not small, approximately 40 degrees. And the perihelion location is nearly 10 au above Neptune's orbit plane. These librations are the result of two orbital resonance conditions: Pluto's mean motion is very close to 1.5 times Neptune's, and its nodal regression rate is equal to its apsidal precession rate (Malhotra & Williams 2000).

These peculiar properties of the then-ninth planet Pluto stimulated several astronomers to theorize about its origins (see, e.g., Marcialis 1997, for a review). Early ideas were that Pluto was an escaped moon of Neptune, escaped possibly during the dynamical capture of Triton from a heliocentric orbit into a retrograde Neptune-centric orbit; but these ideas did not provide a viable explanation for the dynamical properties of Pluto's orbit. Perhaps the first to do this was an idea proposed by this author in a short paper that linked the planetesimal-driven migration of the giant planets to an adiabatic resonance sweeping and capture of Pluto into its orbital resonance with Neptune (Malhotra 1993). As Neptune migrated outward, the locations of Neptune's MMRs also migrated outward. Allowing that Pluto formed in a nearly co-planar, nearly circular orbit somewhere beyond Neptune, it would have been captured in the resonant orbit and swept along when it encountered the sweeping 3/2 MMR. Pluto's capture in resonance would be assured if its initial orbit were of eccentricity below ~0.03 and Neptune's migration rate was sufficiently slow, with an e-folding timescale $\gg 10^5$ yr (see below); the resonance capture would be probable for higher initial eccentricity and the migration rate may also influence the resonance capture probability. After capture, as Neptune continued to migrate, Pluto would remain in libration in the adiabatically migrating resonance. During this migration Pluto's orbital eccentricity would increase in concert with how far Neptune migrated. Malhotra (1993) derived the following relationship between the eccentricity of a resonantly captured minor planet and Neptune's migration:

$$e_{P,final}^2 - e_{P,initial}^2 \approx \frac{1}{j+1} \ln \frac{a_{\text{N,final}}}{a_{\text{N,initial}}} \tag{1}$$

where $e_P$ represents a minor planet's orbital eccentricity, $a_N$ represents Neptune's semimajor axis and $(j+1)$ is the integer describing Neptune's external $(j+1)/j$ MMR with the minor planet (so, $(j+1) = 3$ for the Neptune-Pluto 3/2 resonance). This simple equation leads to the conclusion that Pluto's current eccentricity of ~0.25 could be explained as a consequence of capture into resonance (from an initially circular orbit) when Neptune was closer to the Sun by ~5.1 au. An improved estimate of the adiabatic theory was later made by Yu & Tremaine (1999) who derived that the adiabatic evolution in the 3/2 resonance conserved the following combination of Pluto's eccentricity and semimajor axis:

$$\sqrt{a} \left( 2 - 3\sqrt{1-e^2} \right). \tag{2}$$

This adiabatic invariant also leads to a similar conclusion: that Pluto's current resonance and eccentricity would be explained if it were captured into Neptune's external 3/2 resonance when Neptune was about 5.4 au closer to the Sun than present.

**Planet migration: theoretical predictions**

*Adiabatic resonance sweeping*

The most direct and powerful prediction of the adiabatic resonance sweeping theory is the link between the eccentricities of resonant objects and the value of Neptune's semi-major axis at the time they were captured in resonance, as given by the adiabatic invariant above. This adiabatic invariant was derived in the co-planar approximation for the 3/2 resonance. Extending the adiabatic theory to three dimensions and generalizing it to $(j+k)/j$ MMRs yields the following adiabatic invariant:

$$\sqrt{a} \left[ j - (j+k)\sqrt{1-e^2} \cos i \right], \tag{3}$$

where $i$ is a minor planet's orbital inclination to Neptune's orbit plane (Gomes 2000). This generalization is interesting in that it recognizes that the resonant excitation can be partitioned into both eccentricity and inclination. Taking Pluto's observed inclination of ~17 degrees and its eccentricity of ~0.25, we can conclude that Pluto was captured into the resonance when Neptune's orbit radius was only ~18 au. This places Neptune at least ~12 au closer to the Sun than present. This estimate is a lower bound on Neptune's outward migration because any Kuiper belt objects (KBOs) captured at earlier times could have eccentricity and inclination excited by greater amounts than Pluto's. Indeed, Plutinos have been discovered with higher eccentricities and inclinations than Pluto's, indicating that Neptune may have started out more than ~12 au closer to the Sun. A similar conclusion follows from the population of "Twotinos", the resonant KBOs discovered in Neptune's external 2/1 MMR.

The total mass in planetesimals required to fuel this extent of Neptune's migration has been estimated with numerical simulations to be in the range of 10-50 Earth-masses, and the concurrent energy and angular momentum exchange with the other giant planets causes an inward migration of Jupiter by a few tenths of an astronomical unit and an outward migration of Saturn and Uranus by a few au each (Hahn & Malhotra 1999, 2005).

Many details of the nature of the planets' migration can be expected to be imprinted in the resonant populations of the Kuiper belt, many of which remain to be investigated in detail. Amongst the dynamical properties of the resonant KBOs that are diagnostic of the nature of the giant planets' migration are:

- the population ratios of resonant and non-resonant KBOs,
- the population ratios of Plutinos and Twotinos,
- the population ratios of asymmetric Twotinos,
- the so-called "Kozai" fraction of Plutinos (this refers to those Plutinos that satisfy the second resonance condition, as described for Pluto above),
- the resonance libration amplitude distributions,
- eccentricity-inclination correlations of resonant KBOs,
- eccentricity and inclination distributions of resonant and non-resonant KBOs.

For adiabatic resonance capture from initially nearly co-planar and nearly circular orbits (of eccentricity $e \lesssim \mu_N^{1/3} \approx 0.03$, where and $\mu_N$ is Neptune's mass in units of the solar mass), the timescale of Neptune's migration must be much longer than the resonant libration periods,

$$T_{migration} \gg \mu_N^{-\frac{2}{3}} P_N \approx 10^5 \text{ yr,} \tag{4}$$

where $P_N$ is Neptune's orbital period.

Smooth, adiabatic migration of the planets under planetesimal-driven migration is not assured. (Malhotra 1993) noted that there are at least two ways in which planet migration would be insufficiently smooth for adiabatic resonance capture to be efficient and for the resonant populations to be securely retained for long times:

(A) If Neptune were to have close encounters with large-mass planetesimals, its migration would be punctuated with large kicks (Zhou et al., 2002). Comparing the resonance width with the size of kicks expected from massive planetesimal encounters leads to the conclusion that adiabatic theory prevails when the mass spectrum of the planetesimals that have gravitational scattering encounters with Neptune contains few or no objects more massive than Mars and no more than a few percent is of sizes exceeding ~1000 km (Murray-Clay & Chiang 2006).

(B) If the giant planets were to encounter strong MMRs amongst themselves as they migrated, their mutual resonant perturbations could break the adiabatic theory for resonance capture. This is a stiff condition on the initial orbits and on the path of migration of all the giant planets. Planet-planet resonant encounters have the potential to cause large perturbations to the entire planetary system, including planet-planet scatterings, the ejection of planets, and destruction of most of the solar system. A large number of papers have explored the consequences of planet-planet resonant encounters on the history of the Kuiper belt (see the recent review paper by Dones et al. 2015, and references therein).

*A second channel for populating Neptune's resonances*

Numerical simulations have also revealed an additional channel for populating Neptune's exterior MMRs with KBOs: gravitational scattering followed by "resonance sticking". This works as follows. As Neptune migrates outward by scattering planetesimals, most of the planetesimals undergo repeated scattering in quick succession and are eventually ejected from the solar system by Jupiter. However, some of the outwardly scattered planetesimals have very long return times for a second scattering encounter. This is due to the rare event of scattering into the vicinity of a MMR which allows an object to evolve to a lower eccentricity orbit (higher perihelion distance) along the chaotic layer of the resonance boundary; this phenomenon has become known as "resonance sticking". Subsequent closest approaches to Neptune then occur at larger separations and have weaker perturbing effects. The time to the next strong close encounter can be several gigayears long, comparable to the age of the solar system. This mechanism explains the prominent dynamical structure of the Kuiper belt known as the "scattered disk" (Duncan & Levison 1997). The fraction of the original Kuiper belt that survives to the present day in this scattered disk is estimated to be about 1% (Gomes et al., 2008).

We do not yet have a good theoretical understanding of the resonance sticking phenomenon. Numerical simulations with the full solar system model show that the third dimension may be essential: that in the chaotic resonance boundary layers the evolution to lower eccentricity is correlated with increase in inclination (Lykawka & Mukai 2007). The evolution to lower eccentricity lifts the perihelion distance (since the semi-major axis is locked to the resonance), but lifts it not higher than ~40 au in most cases; the corresponding limit to the inclination excitation is ~40 degrees. This means that the long-lived scattered disk objects are expected to be confined to perihelion distances 30 au $\lesssim q \lesssim$ 40 au (where the lower limit is Neptune's orbit radius) and orbital inclinations up to ~40 degrees.

Numerical simulations also show that the combination of gravitational scattering and resonance sticking is less efficient in populating the resonances than adiabatic resonance capture. It leads to weak capture in resonance with typically large libration amplitudes, in contrast with the adiabatic capture which results in strong resonance capture with small-to-moderate libration amplitudes.

Lykawka & Mukai (2007) and others have noted that, in contrast with the adiabatic resonance capture, for the scattering dynamics the most prominent "sticky resonances" are those in the N/1 sequence (2/1, 3/1, 4/1, …), followed by the N/2 sequence (3/2, 5/2, 7/2, …), and so on. A theoretical explanation for this pattern can be found in the simplified model of the circular, planar restricted three body model of the Sun, Neptune and a test particle (Pan & Sari 2004). With this simplified model, Lan & Malhotra (2019) measured the widths of many of Neptune's stable resonance zones at high eccentricities (Figure 1). We found that in the perihelion distance range 30-35 au, the N/1 sequence of resonances has the largest stable libration zones, followed by the N/2 sequence. Moreover, the widths of the stable libration zones decrease only rather slowly with increasing N, and thereby account for a significant fraction of the area in the semimajor axis-eccentricity parameter range in which gravitationally scattered particles evolve.

This is a rather surprising result, as most previous discussions, including in textbooks on solar system dynamics, have suggested that the overlap of neighboring resonances at planet-crossing eccentricities leads to chaos and complete destruction of the stable libration zones of resonances (e.g., Murray & Dermott, 1999).

As an aside, we note the discovery of a previously unrecognized phase-shifted resonance libration zone at high eccentricities exceeding the Neptune-crossing eccentricity (Lan & Malhotra, 2019). This second libration zone exists for all resonances in the three body model, but in the real solar system it does not support long term stable orbits because most such high eccentricity orbits are not phase-protected from close encounters with Uranus, Saturn and Jupiter. However, this second libration zone may support resonant orbits for short timescales, $\lesssim$ 1 megayears, such as the scattering population of distant Centaurs (Malhotra et al., 2018).

**Long term stability of resonant KBOs**

While the dynamics of resonant KBOs is dominated by the gravitational perturbations of Neptune, the effects of other planets are important over gigayear long timescales. The perturbations from the other giant planets induce weak instabilities and slow chaotic diffusion which cause a gradual erosion of the adiabatically captured resonant populations on gigayear long timescales. Significantly, these erosion rates are different for different resonances: numerical simulations find that the Twotino population erodes faster than the Plutino population; if these two resonances had equal initial populations, four gigayears later the Plutino/Twotino population ratio would be about 2 (Tiscareno & Malhotra 2009).

Even Pluto's gravity affects the long term stability of the Plutino population (Yu & Tremaine 1999). Those Plutinos with eccentricity similar to Pluto's are more stable because they are more likely to have "tadpole" or "horseshoe"-like phases relative to Pluto, analogous to the librations supported by the triangular Lagrange points, L4 and L5, in the classical restricted three body problem. Those Plutinos with eccentricity significantly different than Pluto's also enjoy greater stability as their encounters with Pluto are of higher relative velocity and therefore have smaller perturbing effect. But Plutinos with intermediate eccentricity difference with Pluto's are less stable as they can be driven out of the 3/2 MMR following close encounters with Pluto.

Overall, the erosion rate of the Plutinos induced by the giant planets dominates the erosion rate induced by Pluto. The long-term erosion of the resonant populations likely contributes to the supply of the short period Jupiter family comets in the inner solar system (Levison & Duncan, 1997, Morbidelli 1997, Yu & Tremaine 1999).

**Theory versus observations**

The current observational sample consists of about 2000 KBOs. Their orbital parameters are displayed in Figure 2. This orbital distribution is subject to heavy observational biases because a KBO's on-sky rate of motion and brightness both decrease rapidly with heliocentric distance, which makes the more distant and smaller objects less detectable. Most discovered objects are larger than ~100 km in size, and are closer than 50 au heliocentric distance. Still, with the measured orbital parameters, we can recognize several dynamical classes (e.g., Gladman et al., 2008).

*Classical KBOs*: These are the non-resonant objects, most concentrated in the semimajor axis range 42-47 au. The inner boundary of this range is near the $\nu_8$ apsidal secular resonance which renders circular orbits unstable (Knezevic et al., 1991), and the outer boundary is near Neptune's 2/1 MMR. Despite the observational selection bias against the discovery of more distant objects, the edge of the classical Kuiper belt near ~50 au appears to be quite real (Allen et al., 2001). Two sub-classes are also recognized within the classical KBOs: the *cold classicals* (those with low-eccentricity $e \lesssim 0.1$, and low inclination, $i \lesssim 5$ degrees), and the *hot classicals* (those with higher eccentricities and inclinations). The cold classicals are thought to be the most undisturbed remnants of the primordial Kuiper belt whose orbits have been mildly excited by means of long-term diffusive chaos (Zhou et al. 2007).

*Resonant KBOs*: These objects are found in Neptune's MMRs, most strikingly in the 3/2 resonance (the Plutinos); smaller populations in the 1/1, 2/1, 5/3, 7/4, 5/2, and several other MMRs have been identified. In the semimajor axis−eccentricity plane, the resonant populations present as a vertical concentration over a range of eccentricities with an upper bound corresponding to perihelion distance $q \approx 26$ au; this upper bound is understood to be owed to the destabilizing effects of Uranus.

*Scattered disk objects:* The "scattered disk" is the prominent structure visible in the semi-major axis − eccentricity plane as a curved wing along perihelion distances concentrated in the narrow range 30 au $\lesssim q \lesssim 38$ au and semimajor axes 30 au $\lesssim a \lesssim 1000$ au. Although most of the known scattered disk objects (SDOs) have heliocentric distance currently closer than ~50 au, we infer from their orbital parameters that a vast population exists over heliocentric distances to ~2000 au. Their total population appears to be comparable to or even exceeding the total population of the resonant and classical KBOs.

*Scattering objects:* These are the very high eccentricity non-resonant objects which have perihelion distances below ~26 au and semimajor axes above 30 au. They are so-named because their orbits are unstable on timescales less than 1 megayear as they have close encounters with Neptune. These are a transitional population between the Kuiper belt and the Centaurs/Jupiter family short-period comets.

*Detached objects:* These are the relatively small number of known objects which have semimajor axes $a \gtrsim 50$ au and perihelion distance $q \gtrsim 40$ au. They are so-named because they are thought to originate from the gravitationally scattered population but have been detached from that population by some mechanism that raised their perihelion distance beyond the limits of the scattered disk. Possible mechanisms include: the action of close stellar encounters or tidal torques in the stellar cluster in which the Sun formed (Fernandez & Brunini 2000), the action of massive planetary embryos in the young Kuiper belt (Silsbee & Tremaine 2018), eccentricity-inclination cycles in sweeping MMRs (Gomes et al. 2005) or slow chaotic diffusion in MMRs over gigayear long timescales (Lykawka & Mukai 2007).

It is apparent from Figure 2 that the resonant populations are quite prominent, as is the population of SDOs and the non-resonant population of the classical KBOs. Overall, the resonant KBOs are roughly one-third of the observational sample. Significant resonant populations have been measured in the following MMRs of Neptune (listed in order of increasing semimajor axis): 1/1, 4/3, 3/2, 5/3, 7/4, 2/1, 5/2, 3/1, 4/1. The most prominent is the Plutinos in the 3/2 MMR, with more than 300 objects known. For sizes $\gtrsim 100$ km (absolute magnitude $\lesssim 8.7$), the intrinsic (de-biased) population of the Plutinos is estimated to be about 8000 (Volk et al. 2016). Accounting for the slow erosion of this population over gigayear timescales leads to the conclusion that ~4 gigayears ago it may have exceeded ~27,000. Similar backward-in-time extrapolations can be applied for each resonance.

For the currently known observational sample, the relative intrinsic (de-biased) populations in some of the resonances are displayed in Figure 3; it should be noted that the uncertainties of these debiased estimates are typically ~50% (Volk et al. 2016). We observe that the intrinsic Plutinos/Twotinos population ratio at present is about ~2. Accounting for the slow differential erosion of these populations over ~4 gigayears implies that their populations ~4 gigayears ago would have been of comparable size. This is marginally consistent with the predictions of adiabatic resonance capture which yields the largest capture efficiencies in the 2/1 MMR followed by the 3/2 MMR (Malhotra 1995).

The case of the 5/2 MMR (at $a \approx 55$ au) presents a puzzle. Its presently known population is only 34, but given its greater distance, its intrinsic (de-biased) population is estimated to be 8500, comparable to the Plutino population (Volk et al. 2016). If confirmed, this is inconsistent with adiabatic resonance capture from an initially cold planetesimal disk (Chiang et al. 2003). Moreover, this population has a peculiar eccentricity distribution, with a strong concentration near $e \approx 0.4$. Stimulated by these puzzling observations, we recently investigated the phase space structure of the 5/2 MMR which had not previously been explored in detail (Malhotra et al. 2018). We discovered that the narrow resonance width of the 5/2 MMR at low eccentricities widens dramatically at higher eccentricities, reaching a maximum near $e \approx 0.4$, then narrows again; at eccentricities exceeding $e \approx 0.5$, the perihelion distance is small enough that perturbations from Uranus have a destabilizing effect. Thus, the likely explanation for the peculiar eccentricity distribution of the observed 5/2 resonant KBOs is that the resonance zone is filled in proportion to the width of the stable libration zone as a function of eccentricity.

We also found that the size of the stable libration zone of the 5/2 MMR is comparable to that of the 3/2 MMR and of the 2/1 MMR. This suggests that the similarity of the intrinsic populations in these resonances is related to the sizes of their stable resonance libration zones. This conjecture can be tested in the future as the observational sample size increases and we can measure more reliably the populations of many more MMRs.

However, unlike the case for the 2/1 and 3/2 MMRs, adiabatic resonance sweeping does not provide a compelling mechanism for populating the 5/2 MMR because this third-order resonance has a very narrow neck at low eccentricities which limits the capture probability. Whether direct gravitational scattering can populate this resonance to the observed level remains to be investigated in detail.

An important observation is that the population of non-resonant objects (the classical KBOs) appears to be comparable to or even exceeds the resonant population. This also indicates that the adiabatic resonance sweeping is not the whole story. The other parts of the story are not well constrained yet; they include the following possibilities (see, e.g., Nesvorny 2018 for a review): Neptune's planetesimal-driven outward migration was not smooth, either because the giant planets encountered MMRs with each other or because the scattered planetesimals included significant numbers of large bodies, perhaps even super-Earth-mass planets; the effects of self-stirring and self-gravity of the primordial planetesimal disk; perturbations from planetary-mass objects beyond Neptune that existed at early times, one or more of which may still remain bound in the distant solar system yet-to-be-discovered; external perturbations, such as rare close encounters with passing stars.

**Concluding remarks**

Our empirical knowledge of the Kuiper belt is at an early stage, similar to the stage that our knowledge of the asteroid belt was circa ~1960, prior to the modern era of large telescopic surveys. In the past two decades dedicated surveys for inner solar system asteroids have led to a dramatic increase in the observational sample of the asteroids and correspondingly dramatic increase in asteroid science. We anticipate that in the next decade the observational sample of KBOs will increase by more than an order of magnitude with the advent of the Large Synoptic Survey Telescope which will detect faint, slow moving distant objects over a large fraction of the sky (Ivezic et al., 2008) and will correspondingly stimulate great strides in Kuiper belt science.

Many details of ancient solar system dynamics can be learned from the study of the resonant populations in the Kuiper belt. Theoretically, we understand that there are at least two channels for populating Neptune's exterior resonances: (i) adiabatic sweeping and (ii) the combination of gravitational scattering and resonance sticking. The most heavily populated 3/2 and 2/1 resonances are most easily understood with adiabatic resonance sweeping during the epoch of planetesimal-driven giant planet migration. The theoretical mechanism for this is well understood and it provides our current best quantitative understanding of the overall extent of migration of the giant planets. However, the full story is more complex, with a role for planet-planet resonant encounters and planet-planet scatterings possibly punctuating the migration of the giant planets; these are also likely recorded in the degree of orbital excitation of the Kuiper belt's dynamical structure that survives to the present day. In future studies it would be useful to identify unique or unambiguous measurable signatures of planet-planet resonant encounters, planet-planet scatterings or ejected planets in the distribution of the resonant KBOs; distant unseen planets may also have a measurable effect on the resonant KBOs. Beyond the early dynamical sculpting of the Kuiper belt, resonance sticking is critical for the persistence of the scattered disk objects over the age of the solar system. Weak perturbations and slow chaotic diffusion in the resonances on gigayear long timescales provide the underlying mechanism for the supply of short period comets from the Kuiper belt to the inner solar system. Deeper theoretical studies are needed to understand the phenomena of resonance sticking and chaotic diffusion.

**Abbreviations**

KBO: Kuiper belt object

MMR: mean motion resonance

SDO: scattered disk object

**Declarations**

*Authors contributions*

RM wrote the manuscript. The author read and approved the final manuscript.

*Availability of data and materials*

The data reported in this review paper are from the published literature.

*Competing interests*

The author declares no competing interests.

*Funding*

This work is supported by NASA (grant NNX14AG93G and 80NSSC19K0785) and NSF (grant AST-1824869), and the Marshall Foundation of Tucson, AZ, USA.

*Acknowledgements*

This review is partly based on nearly three decades of research which has greatly benefited from the work of my graduate students Matthew Tiscareno, Kathryn Volk, Xianyu Wang and Lei Lan, as well as many colleagues world-wide. Comments from two anonymous reviewers helped to improve this manuscript. In this brief review I did not attempt to cover the literature comprehensively but have gathered a bibliography that should provide a guide for the interested reader. I thank the AOGS-2018 organizers for inviting this review and for travel support. I am grateful to the US taxpayers for long-term support of astronomical data infrastructure (the Jet Propulsion Laboratory's Horizons service, the Minor Planet Center, and NASA's Astrophysics Data System).

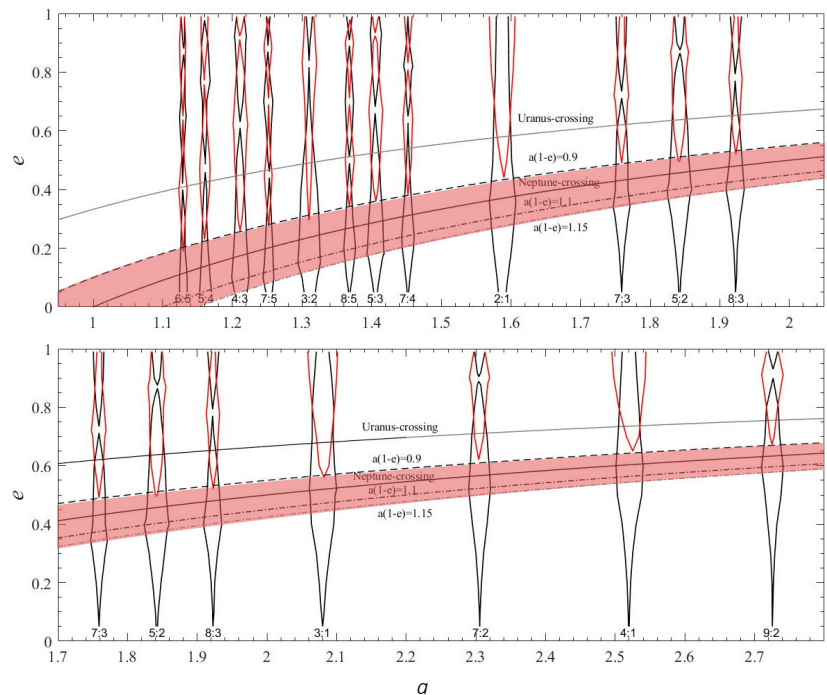

Figure 1: The boundaries of stable libration zones of Neptune's exterior mean motion resonances in the semimajor axis – eccentricity plane; the abscissa is in units of Neptune's semi-major axis. The nearly-vertical curves in black and red indicate the stable libration zones of mean motion resonances; the zones bounded in red represent previously unknown phase-shifted libration zones which exist only at higher eccentricities. The shaded zone bounded by curves of constant

perihelion distance, $q=a(1-e) = 0.9$ to $q = 1.15$ indicates the approximate boundaries of the scattered disk. (Figure adapted from Lan & Malhotra, 2019.)

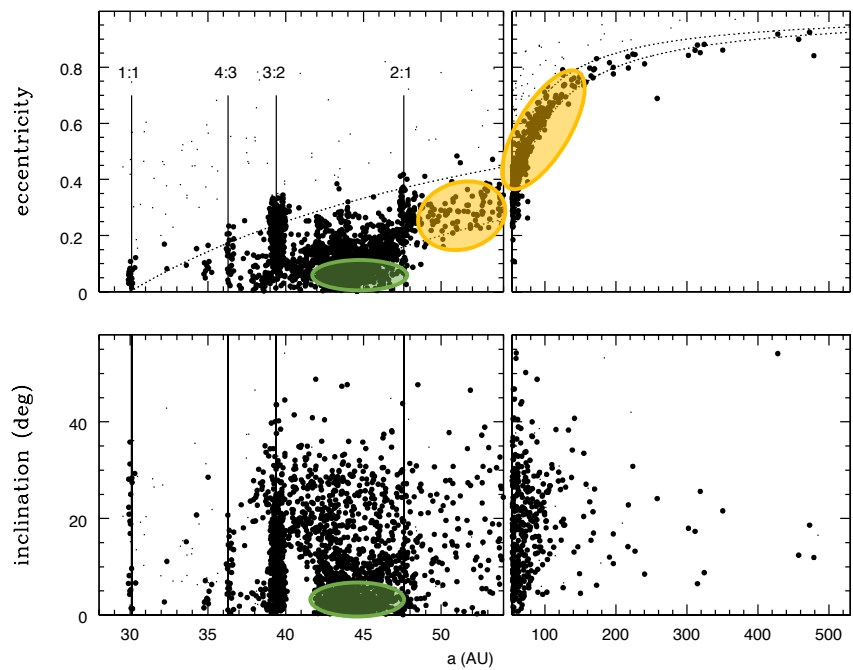

Figure 2: The orbital distribution of known Kuiper belt objects. The vertical lines mark the locations of some prominent mean motion resonances of Neptune. The cold classicals and the scattered disk objects are indicated with green and orange ovals, respectively. (Based on data from the Minor Planet Center, 6 June 2019.)

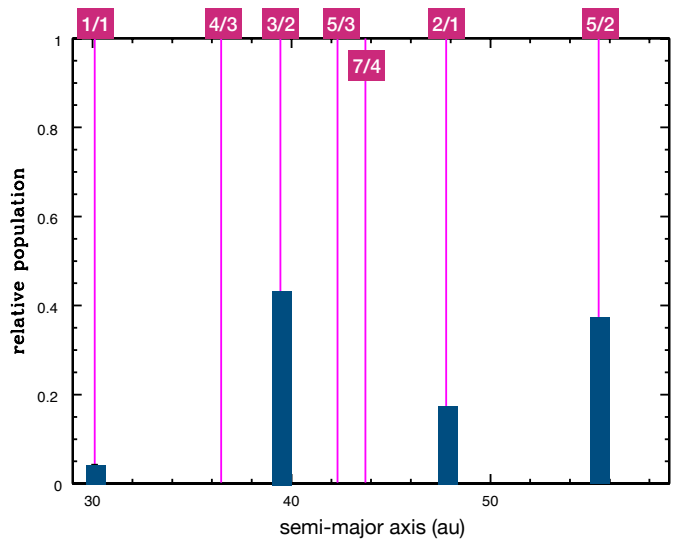

Figure 3: The estimated intrinsic relative populations of resonant Kuiper Belt objects. Only those resonances with published debiased estimates are shown. (Based on data from Volk et al. 2016.)